\documentclass[11pt, a4paper]{article}

\usepackage{ifthen}

\newcommand{\ifndef}[2]{\ifthenelse{\isundefined{#1}}{#2}{}}

\newcommand{\mydef}[2]{\def#1{#2}}

\newcommand{\nospell}[1]{#1}   %

{}
\usepackage{amssymb}
\usepackage{amsmath}   %
{}\usepackage{amsthm}{}   %
\usepackage{latexsym}
\usepackage{amsfonts}
\usepackage{units}     %
\usepackage{psfrag}    %
\usepackage{url}    %
\usepackage{hyphenat}   %
{}
{}
\usepackage[dvips]{graphicx}
\usepackage[usenames,dvipsnames]{color}
{}
{}
{}

{}  %
{}  %
\ifndef{\theorem}{\newtheorem{theorem}{Theorem}[section]}
\ifndef{\lemma}{\newtheorem{lemma}[theorem]{Lemma}}
\ifndef{\corollary}{\newtheorem{corollary}[theorem]{Corollary}}
\ifndef{\conjecture}{\newtheorem{conjecture}[theorem]{Conjecture}}
\ifndef{\remark}{\theoremstyle{remark} \newtheorem{remark}{Remark}}
\ifndef{\proposition}{\newtheorem{proposition}{Proposition}}
\ifndef{\claim}{\newtheorem{claim}[theorem]{Claim}}
\ifndef{\result}{\newtheorem{result}[theorem]{Result}}
\ifndef{\problem}{\newtheorem{problem}{Problem}}

\newtheoremstyle{mydefinition}   %
{\topsep}{\topsep}   %
{\slshape}   %
{}   %
{\bfseries}   %
{.}   %
{ }   %
{}   %
{\theoremstyle{mydefinition}}

\newtheoremstyle{myremark}   %
{\topsep}{\topsep}   %
{\slshape}   %
{}   %
{\bfseries\slshape}   %
{:}   %
{ }   %
{}   %
{\theoremstyle{myremark}}

\newtheoremstyle{myexample}   %
{\topsep}{\topsep}   %
{\itshape}   %
{}   %
{\slshape}   %
{:}   %
{ }   %
{\ul{\thmname{#1}}}   %
\ifndef{\example}{\theoremstyle{myexample} \newtheorem{example}{Example}}

\newtheoremstyle{myclaims}   %
{\topsep}{\topsep}   %
{\slshape}   %
{}   %
{\bfseries\itshape}   %
{}   %
{ }   %
{\thmname{#1}\thmnumber{ \!#2}.}   %
{\theoremstyle{myclaims}

\ifndef{\fact}{\newtheorem{fact}[theorem]{Fact}}
}
{}  %
{}  %

{}
\newtheoremstyle{anystatement}{\topsep}{\topsep}{\itshape}{}{\bfseries}{.}{ }{\anystatementname}
{\theoremstyle{anystatement}}
\newcommand{\anystatementname}{}

{}

\newcommand{\AuxNew}[4][]{#2{#3}[1][*]%
{\ifthenelse{\equal{*}{##1}} %
{\Ensuremath{#1{#4}}}%
{\ifthenelse{\equal{b}{##1}} %
{\Ensuremath{\mathbf{#4}}}%
{\ifthenelse{\equal{}{##1}} %
{\IfMathMode{#1{#4}}{#4}}{}}}}}

\newcommand{\newident}[3][*]{\ifthenelse{\equal{*}{#1}}%
{\AuxNew[\mathit]{\newcommand}{#2}{#3}} %
{\mydef{#2}{\Ensuremath{\mathit{#3}}}}} %

\newcommand{\newidentarg}[2]{%
\newcommand{#1}[1][]%
{\Ensuremath{\mathit{#2}}}}              

\newcommand{\newmat}[3][*]{\ifthenelse{\equal{*}{#1}}%
{\AuxNew{\newcommand}{#2}{#3}} %
{\mydef{#2}{\Ensuremath{#3}}}} %

\newcommand{\providemat}[3][*]{\ifthenelse{\equal{*}{#1}} %
{\AuxNew{\providecommand}{#2}{#3}} %
{\mydef{#2}{\Ensuremath{#3}}}} %

\newcommand{\providematarg}[2]{ %
\providecommand{#1}[1][]{\Ensuremath{#2}}}      

\newcommand{\newmatop}[2]{\mydef{#1}{\operatorname{#2}}}

\newcommand{\newfunction}[2]{ %
\newcommand{#1}[2][*]{\ifthenelse{\equal{*}{##1}}%
{\Ensuremath{#2{\left(##2\right)}}}%
{#2(##2)}}%
}

\newcommand{\MyMakeTheoMacros}[3]{
\newcommand{#2}[2][]{\ifthenelse{\equal{}{##1}}
{\begin{#1} ##2 \end{#1}}
{\begin{#1}\label{##1} ##2\end{#1}}}
\newcommand{#3}[3][]{\ifthenelse{\equal{}{##1}}
{\begin{#1}[##2] ##3 \end{#1}}
{\begin{#1}[##2]\label{##1} ##3\end{#1}}}
}

\makeatletter
\newtheorem*{rep@theorem}{\rep@title}
\newcommand{\newreptheorem}[2]{%
\newenvironment{rep#1}[1]{%
\def\rep@title{#2 \ref{##1}}%
\begin{rep@theorem}}%
{\end{rep@theorem}}}
\makeatother

\newcommand{\MyMakeDupTheoMacros}[7]{
\MyMakeTheoMacros{#1}{#2}{#3}
\newreptheorem{#1}{#6}
\newcommand{#4}[3]{
\newcommand{##2}{##3}
\begin{#1}\label{##1} ##2\end{#1}}
\newcommand{#5}[4]{
\newcommand{##2}{##4}
\begin{#1}{\e{##3}}\label{##1} ##2\end{#1}}
\newcommand{#7}[2]{\begin{rep#1}{##1} ##2 \end{rep#1}}
}

\newcommand{\MyMakeRefMacros}[3]{\newcommand{#1}[2][]
{\ifthenelse{\equal{}{##1}}{#2~\ref{##2}}{#3~\ref{##1} and~\ref{##2}}}}

\newcommand{\MyMakeEqRefMacros}[3]{\newcommand{#1}[2][]
{\ifthenelse{\equal{}{##1}}{#2~\eqref{##2}}{#3~\eqref{##1} and~\eqref{##2}}}}

{}
{}   %
\newcommand{\bibentry}[8]{
{}\bibitem[\nospell{#8}]{#1} {\textup #3}.{}
\ifthenelse{\equal{}{#6}}
{\newblock \textrm{#4.} \newblock {\em #5}, #7.}
{\newblock \textrm{#4.} \newblock {\em #5, #6}, #7.}
}

{}  %
{}  %

\MyMakeTheoMacros{fact}{\fct}{\nfct}

\MyMakeRefMacros{\fctref}{Fact}{Facts}

{}  %
\MyMakeDupTheoMacros{lemma}
{\lem}{\nlem}{\lemdup}{\nlemdup}{Lemma}{\lemrep}
{}

\MyMakeRefMacros{\lemref}{Lemma}{Lemmas}

\MyMakeDupTheoMacros{corollary}
{\crl}{\ncrl}{\crldup}{\ncrldup}{Corollary}{\crlrep}

\MyMakeRefMacros{\crlref}{Corollary}{Corollaries}

\MyMakeTheoMacros{proposition}{\prp}{\nprp}

{}
\newtheorem*{prp*}{\e{Proposition}}

{}

\MyMakeRefMacros{\prpref}{Proposition}{Propositions}

\MyMakeDupTheoMacros{my_claim}
{\clm}{\nclm}{\clmdup}{\nclmdup}{Claim}{\clmrep}

\MyMakeRefMacros{\clmref}{Claim}{Claims}

\MyMakeDupTheoMacros{theorem}
{\theo}{\ntheo}{\theodup}{\ntheodup}{Theorem}{\theorep}

\MyMakeRefMacros{\theoref}{Theorem}{Theorems}

\MyMakeTheoMacros{definition}{\defi}{\ndefi}

\MyMakeRefMacros{\defiref}{Definition}{Definitions}

\MyMakeTheoMacros{problem}{\prob}{\nprob}

\MyMakeRefMacros{\probref}{Problem}{Problems}

\MyMakeTheoMacros{conjecture}{\conj}{\nconj}

\MyMakeRefMacros{\conjref}{Conjecture}{Conjectures}

\providecommand{\qedsymbol}{\square}

\newcommand{\prf}[2][]{\ifthenelse{\equal{}{#1}}%
{\begin{proof}\renewcommand{\qedsymbol}{$\blacksquare$}%
#2 \end{proof}}%
{\begin{proof}[Proof of #1]%
\renewcommand{\qedsymbol}{$\blacksquare_{\mbox{\it{\scriptsize{#1}}}}$}%
#2 \end{proof}}
}

\newcommand{\abstr}[1]{\begin{abstract} #1 \end{abstract}}

\newcommand{\sect}[2][]{\ifthenelse{\equal{}{#1}}
{\section{#2}}
{\section{#2}\label{#1}}}

\newcommand{\ssect}[2][]{\ifthenelse{\equal{}{#1}}
{\subsection{#2}}
{\subsection{#2}\label{#1}}}

\MyMakeRefMacros{\chref}{Chapter}{Chapters}

\MyMakeRefMacros{\sref}{Section}{Sections}

\MyMakeRefMacros{\ssref}{Subsection}{Subsections}

\MyMakeRefMacros{\sssref}{Subsection}{Subsections}

\MyMakeRefMacros{\figref}{Figure}{Figures}

\newcommand{\IfMathMode}[2]{\ifmmode{#1}\else{#2}\fi}

\newcommand{\Ensuremath}{\ensuremath}

\newcommand{\fbr}[1]{\IfMathMode %
{#1}{$#1$}}                      %

\newcommand{\fnbr}[1]{\mbox{\fbr{#1}}}   %

\newcommand{\fla}[2][*]{\ifthenelse{\equal{}{#1}}{\fbr{#2}}{\fnbr{#2}}}

\newcommand{\mat}[2][]{\ifthenelse{\equal{}{#1}} %
{ \begin{displaymath} #2 \end{displaymath} } %
{ \begin{equation} \label{#1} #2 \end{equation} }%
}

\newcommand{\f}{\fla}

\newcommand{\m}{\mat}

\MyMakeEqRefMacros{\equref}{Equation}{Equations}

\MyMakeEqRefMacros{\expref}{Expression}{Expressions}

\MyMakeEqRefMacros{\inequref}{Inequality}{Inequalities}

\newcommand{\bracref}[1]{(\ref{#1})}

\newcommand{\bref}{\bracref}

\providecommand{\middle}{\big}

\newcommand{\chs}{\genfrac(){0cm}{}}   %

\newmatop{\rank}{rank}

\providecommand{\E}[2][]{\ifthenelse{\equal{}{#1}}%
{\mathop{\mathbf{E}}{\left[{#2}\right]}}%
{\mathop{\mathbf{E}}_{#1}{\left[{#2}\right]}}}

\newcommand{\PR}[2][]{\mathop{\mathbf{Pr}}_{#1}{\left[{#2}\right]}}

\newcommand{\U}[1][]{\ifthenelse{\equal{}{#1}}%
{{\cal U}}%
{{\cal U}_{#1}}}

\newcommand{\fr}[3][*]{%
\ifthenelse{\equal{*}{#1}}        %
{\frac{#2}{#3}}{}%
\ifthenelse{\equal{/}{#1}}        %
{\nicefrac{#2}{#3}}{}%
\ifthenelse{\equal{}{#1}}         %
{\left.#2\middle/#3\right.}{}%
\ifthenelse{\equal{p_}{#1}}       %
{\left.\left(#2\right)\middle/#3\right.}{}%
\ifthenelse{\equal{_p}{#1}}       %
{\left.#2\middle/\left(#3\right)\right.}{}%
\ifthenelse{\equal{pp}{#1}}       %
{\left.\left(#2\right)\middle/\left(#3\right)\right.}{}
}

\newcommand{\dr}{\nicefrac}

\def\MySQRT#1#2{    %
\setbox0=\hbox{$#1\sqrt{#2\,}$}\dimen0=\ht0%
\advance\dimen0-0.2\ht0%
\setbox2=\hbox{\vrule height\ht0 depth -\dimen0}%
{\box0\lower0.4pt\box2}}

\newcommand{\set}[2][]{\ifthenelse{\equal{}{#1}} %
{\Ensuremath{\left\{#2\right\}}}%
{\Ensuremath{\left\{#2\,\middle\arrowvert\,#1\right\}}}}

\newcommand{\Minn}[3][]{\ifthenelse{\equal{}{#1}} %
{\Ensuremath{\min_{#2}{\left\{#3\right\}}}}%
{\Ensuremath{\min_{#2}{\left\{#3\,\middle\arrowvert\,#1\right\}}}}}

\newfunction{\asO}{O}
\newfunction{\aso}{o}
\newfunction{\asOm}{\Omega}
\newfunction{\asT}{\Theta}

\mydef{\01}{\set{0,1}}

\renewcommand{\l}{\left}
\renewcommand{\r}{\right}

\newcommand{\sz}[2][]{\ifthenelse{\equal{}{#1}}%
{\Ensuremath{\left|#2\right|}}%
{\Ensuremath{\left|#2\middle|_{#1}\right.}}}
\providecommand{\norm}[2][]{\ifthenelse{\equal{}{#1}}%
{\Ensuremath{\left\|#2\right\|}}%
{\Ensuremath{\left\|#2\middle\|_{#1}\right.}}}

\newcommand{\txt}[1]{\textrm{#1}}   %

\newcommand{\Cl}[1]{{\cal #1}} %

\DeclareMathAlphabet{\lowcal}{OT1}{pzc}{m}{it}

\newidentarg{\RI}{\mathcal R_{#1}^{1}}
\newidentarg{\RII}{\mathcal R_{#1}^{\parallel}}
\newidentarg{\RIIp}{\mathcal R_{#1}^{\parallel,pub}}
\newidentarg{\Rz}{\mathcal R_{#1}^{\leftrightarrow}}

\newmat[]{\lla}{\left\langle}
\newmat[]{\rra}{\right\rangle}

\newmat[]{\LRarr}{\Leftrightarrow}

\newmat[]{\tm}{\cdot}
\newmat[]{\sbseq}{\subseteq}
\newmat[]{\sbs}{\subset}
\newmat[]{\smin}{\setminus}
\newmat[]{\eps}{\varepsilon}
\newmat[]{\deq}{\stackrel{\textrm{def}}{=}}

\providemat{\QQ}{\mathbb{Q}}
\providematarg{\NN}{\ifthenelse{\equal{}{#1}}%
{\mathbb{N}}%
{\mathbb{N}_{#1}}}

\newcommand{\ds}[1][]
{\ifthenelse{\equal{}{#1}}{\dots}{#1\dots#1}}
\newmat[]{\dc}{\ds[,]}

\newcommand{\enum}[1]{\begin{enumerate} #1 \end{enumerate}}

\newcommand{\itemi}[2][]{\ifthenelse{\equal{}{#1}}
{\begin{itemize} #2 \end{itemize}}
{\begin{itemize}[#1] #2 \end{itemize}}}

\newcommand{\bul}[1]{\itemi{\item #1}}

\newcommand{\MyComment}[1]{\ClassWarning{My Macros}{#1}}

\newcommand{\fn}{\footnote}

{}  %
\newcommand{\e}{\emph}
{}   %

\providecommand{\ul}[1]{\underline{#1}}  %

{}
{}

\newmatop{\dst}{d_{st}}

\title{Partition Expanders}

\author{Dmitry Gavinsky%
\thanks{Institute of Mathematics, Academy of Sciences, \v Zitna 25, Praha 1, Czech Republic.
Partially funded by the grant P202/12/G061 of GA \v CR and by RVO:\ 67985840.}
\and Pavel Pudl\'ak\footnotemark[1]
}

\date{}

\begin{document}

\maketitle

\thispagestyle{empty}

\abstr{We introduce a new concept, which we call \emph{partition
expanders}. The basic idea is to study quantitative properties of
graphs in a slightly different way than it is in the standard
definition of expanders. While in the definition of expanders it is
required that the number of edges between any pair of sufficiently
large sets is close to the expected number, we
consider \emph{partitions} and require this condition only for \emph{most
of the pairs} of blocks. As a result, the blocks can be substantially
smaller.

We show that for some range of parameters, to be a partition expander a random graph needs \e{exponentially smaller} degree than any expander would require in order to achieve similar expanding properties.

We apply the concept of partition expanders in communication complexity.
First, we give a PRG for the SMP model of the optimal seed length, $n+\asO{\log k}$.
Second, we compare the model of SMP to that of Simultaneous Two-Way Communication, and give a new separation that is stronger both qualitatively and quantitatively than the previously known ones.}

\setcounter{page}{1}

\sect{Introduction}

Expanders are a very interesting and useful concept and appear in many
applications in computer science. Therefore several related concepts
have been introduced; e.g., lossless expanders \cite{CRVW02_Ran}, monotone expanders and dimension expanders \cite{DW10_Mon}, superexpanders \cite{MN13_Non}.

In this paper we introduce yet another concept that we call \emph{partition expanders}. The definition is motivated by the following observation.
The well-known Expander-Mixing Lemma says, roughly
speaking, that for every two sufficiently big sets of vertices $A$ and
$B$ the number of edges of the expander between $A$ and $B$ is close
to $\fr dn\tm |A|\tm|B|$, where $n$ is the number of vertices and $d$
is the degree. If we want to apply this lemma to smaller sets, we have
to increase the degree of expanders appropriately.

Now suppose we have
a partition of the vertices of the graph and we only want to satisfy the
density condition for most of the pairs of sets. 
It turns out that a random graph
with relatively small degree is able to satisfy this condition for
partitions with many blocks, although the Expander-Mixing Lemma is not
able to give any interesting estimate. So while expanders are graphs
with ``typical connectivity'' with respect to \e{subsets of vertices},
partition expanders have ``typical connectivity'' with respect to
\e{partitions of vertices}.
Informally speaking, in the context of expanders, partitions are ``more structured'' objects than subsets, and therefore demanding the same ``expanding performance'' with respect to partitions
can be viewed as a relaxation of usual expanders.
In return, we expect partition expanders to have considerably smaller degree than usual expanders with the same expanding performance.

There are several possible ways to formally define a partition
expander.  We choose the following definition as ``canonical'' due to
its brevity and robustness. We will give alternative definitions shortly.

\ndefi[d_pex]{Partition expanders}{Let $G=(V,E)$ be an (undirected)
graph.  Let $\mu$ be the uniform distribution over $V\times V$, and
let $\mu_G$ be the uniform distribution over $E$.  For any coloring
$c:V\to[K]$, let $\nu^c$ and $\nu_G^c$ be the distributions of the
pair $(c(v_1),c(v_2))$ when $(v_1, v_2)$ is chosen according to
$\mu$ or $\mu_G$, respectively.

For $K\in\NN$ and $\delta\in(0,1)$, we say that $G$ is a
$(K,\delta)$-\emph{partition expander} if for every coloring $c:V\to[K]$ the
statistical distance between $\nu^c$ and $\nu_G^c$ is at most
$\delta$.}

It should be noted that this concept is interesting in the situations
where the number $K$ of partitions is increasing with the number of
vertices and the graphs are $d$-regular with $d$ increasing.
We are mainly interested in the question of how
small $d$ can be for a given $K$, assuming $0<\delta<1$ is a fixed constant.

\ssect{Our results}

We start by giving several equivalent definitions of partition expanders, which emphasize the fact that they are a natural modification of usual expanders. 

In Section~\ref{s_rand} we analyze the behavior of random graphs as partition expanders.
We prove that random $d$-regular graphs almost always are good partition expanders -- the dependence of $K$ on $d$ is the best possible, namely exponential.

In Section~\ref{s_peve} the notion of partition expanders is advocated through comparing it to expanders.
We show that the gap between the absolute values of the first two eigenvalues does not ensure that the graph is a good partition expander.
Namely, if only the spectral gap is taken into account when a partition expander is constructed, then the degree has to be \e{exponentially larger} than an optimal partition expander requires.
Since the spectral gap characterizes almost tightly the expander properties of a graph, this demonstrates exponential advantage of partition expanders (in those scenarios when they are suitable) over expanders.
In other words, if ``partition expansion'' is the desired behavior, then using an expander instead of an optimal partition expander would incur exponential loss in terms of the required degree.

Based on the spectral properties only, we use the Hoffman-Wielandt
inequality and get a slightly better bound than what would follow from
a direct application of the Expander-Mixing Lemma.\fn
{We get quadratic improvement in terms of the partition size, and show that it is essentially optimal general bound in terms of the spectral gap alone.}
The fact that the spectral gap is incapable to characterize good partition expanders partially explains why new methods are required for their construction.

In \sref{s_SMP} we present another equivalent definition of partition
expanders.
We show that a graph $G=(V,E)$ is a partition expander if and only if the uniform distribution over $E$ is a \e{Pseudo-Random Generator (PRG)} in the setting of \e{Simultaneous Message Passing (SMP)} in communication complexity.  
We use this fact to give a lower bound on the degree of partition expanders, thus showing \e{optimality} of the randomized construction given in \sref{s_rand}.

In the second part of \sref{s_SMP} we show two applications of our randomized construction of a partition expander.
First, we construct a PRG against SMP protocols of communication cost $k$ that requires seed length $n+\asO{\log k}$ (see \theoref{t_equi2} and the comment thereafter).\fn
{All previously known PRGs in communication complexity were given against stronger models, thus requiring exponentially larger ``overhead'' over $n$ in terms of seed length -- for details, see \sref{s_SMP}.}
Second, we compare the model of SMP to that of Simultaneous Two-Way Communication, and give a new separation that is stronger both qualitatively and quantitatively than the previously known ones (see \theoref{t_sep}).

\sect[s_not]{Notation and more}

Unless stated otherwise, all sets are assumed to be finite, and all graphs are \e{undirected} and \e{simple} (having no self loops or multiple edges).\fn
{In those cases when we explicitly allow multiple edges, the edges of a graph will be viewed as a collection with repetitions.}  
For two subsets $S_1,S_2\sbseq V$, we denote by
$E(S_1,S_2)$ the set of ordered pairs $(v_1,v_2)$ such that  $(v_1,v_2)$ is an
edge in $E$, 
$v_1\in S_1$ and $v_2\in S_2$, and write $E(v_1,v_2)$ for
$E(\set{v_1},\set{v_2})$.%
\footnote{Note that if $v_1,v_2\in S_1\cap S_2$, then the edge
$(v_1,v_2)$ appears in $E(S_1,S_2)$ twice: as ordered pairs  $(v_1,v_2)$ and
$(v_2,v_1)$.}
We will say that a set family $\sigma=\set{C_1\dc C_K}$ is a \f K-partition of a set $X$ if $\cup_{i=1}^KC_i=X$ and $C_1\dc C_K$ are pairwise disjoint and nonempty.

The \e{statistical distance} between two distributions $\mu_1$ and $\mu_2$
defined over a set $X$ is 
$$\dst(\mu_1,\mu_2)\deq\fr12\sum_{x\in X}\sz{\mu_1(x)-\mu_2(x)}.$$

\lem[l_equi]{Let $K\in\NN$ and $\delta\in\mathbb{R}$.
The following statements are equivalent:\enum{
\item \label{it_pe} $G=(V,E)$ is a $(K,\delta)$-partition expander.
\item \label{it_sig} For every $K$-partition $\sigma=\set{C_1\dc C_K}$ of $V$,
\begin{equation}\label{e1}
\delta\ge\fr12\sum_{i,j\in[K]}
\sz{\fr{\sz{E(C_{i},C_{j})}}{\sz E}
-\fr{\sz{C_{i}}\tm\sz{C_{j}}}{\sz V^2}}.
\end{equation}
\item \label{it_sig-S} For every $K$-partition $\sigma$ and
$S\sbseq[K]\times[K]$, 
\begin{equation}\label{e1.5}
{\delta\ge\sum_{(i,j)\in S}
\l(\fr{\sz{E(C_{i},C_{j})}}{\sz E}
-\fr{\sz{C_{i}}\tm\sz{C_{j}}}{\sz V^2}\r)
=\fr{\sum_S\sz{E(C_{i},C_{j})}}{\sz E}
-\fr{\sum_S\sz{C_{i}}\tm\sz{C_{j}}}{\sz V^2}.}
\end{equation}
\item \label{it_sig-S-sym} Like \ref{it_sig-S}, but only over symmetric $S$ (i.e., $(i,j)\in S\LRarr(j,i)\in S$).
}}

\prf{Equivalence between \ref{it_pe} and \ref{it_sig} is immediate from \defiref{d_pex}.
Equivalence between \ref{it_sig-S} and \ref{it_sig-S-sym} follows from the fact that $G$ is undirected.
To see that \ref{it_sig} is equivalent to \ref{it_sig-S}, note that
\m{
\sum_{i,j\in[K]}\l(\fr{\sz{E(C_{i},C_{j})}}{\sz E}
-\fr{\sz{C_{i}}\tm\sz{C_{j}}}{\sz V^2}\r)
=\fr{\sum_{[K]\times[K]}\sz{E(C_{i},C_{j})}}{\sz E}
-\fr{\sum_{[K]\times[K]}\sz{C_{i}}\tm\sz{C_{j}}}{\sz V^2}\\
=0.}
}

There are many possible ways to define expanders. The standard
definition is based on the second largest absolute value of an
eigenvalue of a graph $G$, which we will denote by $\lambda(G)$. 

\ndefi{Expanders}{A regular graph $G$ is an $\ell$-expander if $\lambda(G)\leq \ell$.}

We will denote the degree of a regular graph $G$ by $d(G)$, or simply by $d$ when $G$ is clear from the context.

The most natural relation between expanders and partition expanders comes from the following well-known fact (e.g., see~\cite{AS08_The}). 

\nlem[l_ExMi]{Expander-Mixing Lemma}{Let $(V,E)$ be an $\ell$-expander. Then for every $S_1,S_2\sbseq V$,
\m{\sz{\fr{\sz{E(S_1,S_2)}}{\sz E}-\fr{\sz{S_1}\tm\sz{S_2}}{\sz V^2}}
\le \ell \tm\fr{\sqrt{\sz{S_1}\tm\sz{S_2}}}{\sz E}=
\frac \ell d \tm\fr{\sqrt{\sz{S_1}\tm\sz{S_2}}}{\sz V}
.}
}

One can show using this lemma that an $\ell$-expander is a
$(K,\delta)$-partition expander for constant $\delta>0$ and certain $K\in\asT{d/\ell}$ -- however, this trivial arguments fails for $K\ge d/\ell$.
In \sref{s_peve} we will use the Hoffman-Wielandt inequality to show that an $\ell$-expander is a $(K,\asOm 1)$-partition expander for certain $K\in\asT{(d/\ell)^2}$, and that will be shown to be optimal up to the factor of $\log n$.

\ntheo[t_HW]{Hoffman-Wielandt inequality~\cite{HW53_The}}{If $A$ and $B$ are normal matrices with respective eigenvalues $\lambda_1(A)\dc\lambda_n(A)$ and $\lambda_1(B)\dc\lambda_n(B)$, then
\m{\Minn{\pi}{\sum_{i=1}^n\sz{\lambda_i(A)-\lambda_{\pi(i)}(B)}^2}\le
\norm{A-B}_F^2,}
where $\pi$ runs over all permutations over $[n]$ and $\|\dots\|^2_F$
denotes the square of the Frobenius norm (the sum of squares of the absolute values of the elements).}

If $A$ and $B$ are symmetric real matrices, we can drop the absolute
value and write the terms as 
$\lambda_i(A)^2+\lambda_{\pi(i)}(B)^2-2\lambda_i(A)\lambda_{\pi(i)}(B)$. Since
the sum of the squares of eigenvalues of a matrix is the square of its Frobenius norm,
the inequality is equivalent to
\begin{equation}\label{e-HW}
\sum_{i,j}a_{ij}b_{ij}\leq\max_\pi\left\{\sum_{i=1}^n\lambda_i(A)\lambda_{\pi(i)}(B)\right\}.
\end{equation}

\bigskip
To prove the existence of good $d$-regular partition expanders, we will use the following well-known bound on the concentration of probability measures.
Recall that a sequence $X_0\dc X_n$ of real-valued random variables is called a martingale if for every $0\leq i<n$, 
${\bf E}[X_{i+1}|X_i]=X_i$. 

\ntheo[t_Azu]{Azuma Inequality~\cite{A67_We}}{Let $X_0\dc X_n$ be
a martingale satisfying 
\m{\forall i\in[n]:\sz{X_i-X_{i-1}}\le c,}
for some real $c>0$. 
Then for any real $t>0$,
\m{\PR{X_n>X_0+t},~\PR{X_n<X_0-t}
\le\exp\l(-\fr{t^2}{2nc^2}\r).}}

A typical situation in which this theorem is applied is when $X_0\dc
X_n$ is a \emph{Doob martingale}, i.e., it is defined using $n$
random variables $Y_1,\dots,Y_n$ and a function $f(y_1,\dots,y_n)$ as
follows:
\[
X_i={\bf E}_{Y_{i+1}\dc Y_n}[f(Y_1\dc Y_{i},Y_{i+1}\dc Y_n)],
\]
for $i=0\dc n$; 
in particular, $X_0={\bf E}[f(Y_1\dc Y_n)]$ and $X_n=f(Y_1\dc Y_n)$.
(One can easily check that this formula defines a
martingale.) Hence we have:

\crl[cr_Azu]{For $m\in\NN$, let $Y_1\dc Y_m$ be random real variables
and let $f:\mathbb{R}^m\to\mathbb{R}$.
Let  $X_0\dc X_n$ be defined as above. 
Then for any real $t>0$,
\m{\PR{f(Y_1\dc Y_m)>\nu+t},~
\PR{f(Y_1\dc Y_m)<\nu-t}
\le\exp\l(-\fr{t^2}{2nc^2}\r),}
where $\nu\deq\E{f(Y_1\dc Y_m)}\ (=X_0)$ and $c$ satisfies $\sz{X_i-X_{i-1}}\le c$.}

Let $d,n\in\NN$ be such that $2|dn$, denote by $\Cl G_{n,d}$ the
uniform distribution on $d$-regular (simple undirected) graphs on $n$
vertices.  In our analysis we will use the \e{pairing method} for
generating $G\sim\Cl G_{n,d}$, due to Bollob\'as~\cite{B80_A_Pro}
(also see~\cite{W99_Mo}).  

\nlem[c_PairMe]{Pairing method~\cite{B80_A_Pro}}{The
following procedure generates $E\sbseq[n]\times[n]$ such that
$G=([n],E)\sim\Cl G_{n,d}$.  
\enum{
\item \label{it_beg} Let $\pi\sbs[nd]\times[nd]$ be a uniformly random perfect matching on $[nd]$ (viewed as a symmetric set of directed edges).
For $i\in[n]$, let $cell_i\deq\set[id-d<x\le id]x$ and $d_\pi(v_1, v_2)\deq\sz{\pi(cell_{v_1},cell_{v_2})}$.
\item  For every $(v_1, v_2)\in[n]\times[n]$, let $(v_1,v_2)$ be $d_\pi(v_1, v_2)$ times an element of $E$.
\item \label{it_Ret} Return to Step \ref{it_beg} if $G=([n],E)$ is not simple.
}}

In the analysis we will consider the distribution of $([n],E)$
resulting from dropping Step~\ref{it_Ret} off the above procedure; let
us denote it by $\Cl G_{n,d}'$.  Observe that a graph $G\sim\Cl
G_{n,d}'$ is always undirected, but doesn't have to be simple.\fn {Note
also that the distribution $\Cl G_{n,d}'$ is not uniform over its
support - e.g., $\Cl G_{2,2}'$ produces the graph with two parallel
edges with probability $2/3$.}

We will use the following estimate, due to McKay and Wormald~\cite{KW91_Asy}:
\nlem[l_simple]{\cite{KW91_Asy}}{For $d\in\aso{\sqrt n}$,
\m{\PR[G\sim\Cl G_{n,d}']{G\txt{ is simple}}\in
\exp\l(\fr{1-d^2}4-\fr{d^3}{12n}+\asO{\fr{d^2}n}\r)\sbseq\exp(\aso n).}
}

\sect[s_rand]{Random $d$-regular graphs as partition expanders}

Let us see that a random regular graph is likely to form a partition expander.

\theo[t_r-reg]{For $d\in\asO{n^{1/3}}$, a random $d$-regular simple
undirected graph on $n$ vertices is a $(K,\delta)$-partition
expander with probability at least $1-\exp\l(n\log
K+K^2-\asOm{\delta^2nd}\r)$.}

\crl[cr_r-reg]{For any $\varepsilon>0$ and $B\in\NN$ there exists $C\in\NN$,
such that the following holds: A random \f d-regular graph on $n$
vertices is a $(K,\delta)$-partition expander with probability at
least $1-\varepsilon$, as long as $K\le B\tm\sqrt n$ and $d\ge\fr{C\tm\log
K}{\delta^2}$.}

To prove the theorem, we will use the following lemma.

\lem[l_Gpi]{Let $2|n$ and $G=([n],E)$ be a simple undirected graph.
Let $\Cl M_n$ be the uniform distribution of perfect matchings on $[n]$.
A universal constant $C$ exists, such that for every $\delta>0$,
\m{\PR[\pi\sim\Cl M_n]{\sz{\pi\cap E}\ge\fr{|E|}{n-1}+\delta n}
<\exp\left(-\fr{\delta^2 n}C\right).}}

Note that $\E{\sz{\pi\cap E}}=\fr{|E|}{n-1}$, and therefore the lemma is a natural tail bound.

\prf[\lemref{l_Gpi}]{Let $m=n/2$.
Selecting $\pi\sim\Cl M_n$ can be achieved via repeating the step
\bul{Let $v_i\sim\U[{[n]\smin\set{v_1\dc v_{i-1}}}]$}
for $i$ running from $1$ to $n$, followed by setting
\m{\pi\deq\bigcup_{i=1}^{m}
\set{(v_{2i-1},v_{2i}),(v_{2i},v_{2i-1})}.}
Let $e_i\deq(v_{2i-1},v_{2i})$, and $s(e)=2$ if $e\in E$ and $s(e)=0$ otherwise.\fn
{Note that if $e=(v_{2i-1},v_{2i})\in E$ then $(v_{2i},v_{2i-1})\in E$ as well.}
Let $S=\sum_{i=1}^ms(e_i)$ (thus $S=|\pi\cap E|$).
Define $X_0, X_1\dc X_m$ by
\m{X_i={\bf E}[S|e_1,\dots,e_i].}
According to \crlref{cr_Azu}, to prove the lemma we only need to show $|X_{i+1}-X_i|\in\asO1$.

Let $0\leq i<m$ and suppose that the vertices $v_1\dc v_{2i}$ have been selected.
Let $k$ be the number of remaining edges, i.e.,
$k=|E\cap([n]\smin\set{v_1\dc v_{2i}})^2|$.
Then
\[
X_i=\sum_{j=1}^is(e_j)+2(m-i)\tm\fr k{\chs{n-2i}2}=
\sum_{j=1}^is(e_j)+\fr{2k}{n-2i-1}.
\]

First suppose that $i<m-1$, and let $k'$ be the number of the remaining edges after $e_{i+1}=(v_{2i+1},v_{2i+2})$ has been selected.
Then
\m{X_{i+1}=\sum_{j=1}^is(e_j)+s(e_{i+1})+\frac {2k'}{n-2i-3},}
and
\m{X_{i+1}-X_i=s(e_{i+1})+\fr{2k'}{n-2i-3}-\fr{2k}{n-2i-1}.}
On the one hand,
\m{X_{i+1}-X_i
<2+k\tm\l(\fr2{n-2i-3}-\fr2{n-2i-1}\r)
\le2+\fr{4(n-2i)(n-2i-1)}{(n-2i-1)(n-2i-3)}
\in\asO1.}
On the other hand,
\m{X_{i+1}-X_i
\ge\fr{2(k'-k)}{n-2i-3}
>-\fr{4(n-2i-2)+1}{n-2i-3}
\in\asO1.}

The case of $i=m-1$ can be treated similarly, and the result follows.}

We are ready to prove the main result of this section.

\prf[\theoref{t_r-reg}]{Let the graph $G=([n],E)$ be sampled from $\Cl G_{n,d}$.
Consider an arbitrary $K$-partition $\sigma=\set{C_1\dc C_K}$ of $[n]$ and a symmetric set $X_+\sbseq[K]\times[K]$, and let us bound the probability of the event
\m{\Cl E(\sigma,X_+)\deq\lla
\fr{\sum_{(i_1,i_2)\in X_+}\sz{E(C_{i_1},C_{i_2})}}{nd}
>\fr{\sum_{(i_1,i_2)\in X_+}
\sz{C_{i_1}}\tm\sz{C_{i_2}}}{n^2}+\delta\rra.}

Assume that the pairing method (\clmref{c_PairMe}) was used to generate $G\sim\Cl G_{n,d}$.
In order to analyze $\PR[\Cl G_{n,d}]{\Cl E(\sigma,X_+)}$, we first look at the corresponding event for $G'=([n],E')\sim\Cl G_{n,d}'$, namely:
\m{\Cl E'(\sigma,X_+)\deq\lla
\fr{\sum_{(i_1,i_2)\in X_+}\sz{\pi(S_{i_1},S_{i_2})}}{nd}
>\fr{\sum_{(i_1,i_2)\in X_+}
\sz{S_{i_1}}\tm\sz{S_{i_2}}}{(nd)^2}+\delta\rra,}
where $S_i\deq\cup_{x\in C_i}cell_x$ and $\pi$ is the perfect matching that has been used for producing $E'$.
From the construction and \lemref{l_simple} it follows that
\m[m_GG']{\PR[\Cl G_{n,d}]{\Cl E(\sigma,X_+)}
\le\frac{\PR[\Cl G_{n,d}']{\Cl E'(\sigma,X_+)}}
{\PR[\Cl G_{n,d}']{G\txt{ is simple}}}
\in\PR[\Cl G_{n,d}']{\Cl E'(\sigma,X_+)}
\tm\exp(o(n)).}
Let
\m{E_{\sigma,X_+}
\deq\cup_{(i_1,i_2)\in X_+}S_{i_1}\times S_{i_2}
\smin\set[{v\in[nd]}]{(v,v)},}

and note that $([nd],E_{\sigma,X_+})$ is a simple graph.
Then
\m{\Cl E'(\sigma,X_+)=\lla\sz{\pi\cap E_{\sigma,X_+}}
>\fr{\sz{E_{\sigma,X_+}}}{nd}+\delta nd\rra.}

By \lemref{l_Gpi},
\m{\PR[\Cl G_{n,d}']{\Cl E'(\sigma,X_+)}=\PR[\pi]{\sz{\pi\cap E_{\sigma,X_+}}
>\fr{\sz{E_{\sigma,X_+}}}{nd}+\delta nd}
\in\exp\l(-\asOm{\delta^2nd}\r).}
From \bref{m_GG'} and our assumption about $d$,
\m{\PR[\Cl G_{n,d}]{\Cl E(\sigma,X_+)}
\in\exp\l(o(n)-\asOm{\delta^2nd}\r)
=\exp\l(-\asOm{\delta^2nd}\r).}
By \lemref{l_equi}, $G$ is not a $(K,\delta)$-partition expander if an only if for some $K$-partition $\sigma$ and a symmetric set $X_+\sbseq[K]\times[K]$, the event $\Cl E(\sigma,X_+)$ holds.
By the union bound, this probability is at most
\m{\exp\l(n\log K+K^2-\asOm{\delta^2nd}\r),}
as required.}

\sect[s_peve]{Partition expanders vs.\ expanders}

Let us compare the notions of expanders and partition expanders in more detail. 

\begin{theorem}\label{t_epe}
Let $G$ be a $d$-regular $\ell$-expander on $n$ vertices.
Then it is a $(K,\sqrt{K}\ell/d)$-partition expander for every $K<d^2/\ell^2$.
\end{theorem}

Note that the Expander-Mixing Lemma (\lemref{l_ExMi}) only gives that $G$ is a
$(K,\delta)$ partition expander for $\delta=O(K\ell/d)$, which is meaningful only for $K<d/\ell$.
The statement of the above theorem is essentially tight (cf.~\theoref{t_HWtight}), and this means that only small (quadratic, in terms of $K$ vs.\ $d$) improvement can result from using partition expanders instead of expanders, as long as the construction of a partition expanders relies on the spectral gap.
On the other hand, we will see soon that good partition expanders offer \e{exponential} improvement in terms of the dependence of $K$ on $d$.

\begin{proof}
Let a $d$-regular graph $G$ on $[n]$ be given. Let $E$ be its
adjacency matrix.  We will use the equivalent definition of
partition expanders from \lemref{l_equi}-\bref{it_sig-S-sym} based
on symmetric sets~$S$.  Let a $K$-partition $\{C_1,\dots,C_K\}$ be
given and let $S\subseteq [K]\times[K]$ be a symmetric set. Let $A$
be the adjacency matrix of the graph that $S$ induces on $[n]$;
i.e.,
\[
A_{ij}=1\ \mbox{ if } i\in C_k,\ j\in C_l \mbox{ for some } (k,l)\in S,
\mbox{ and } =0\ \mbox{ otherwise.}
\]
Note that $\rank(A)\le K$.

Sampling uniformly from the edges of $G$ is represented by the matrix
$\frac 1{nd}E$.  Sampling uniformly from all the edges is represented
by the matrix $\frac 1{n^2}J$, where $J$ is the matrix of 1s.  Hence
we need to bound the scalar product of the matrices $A$ and $B$
(viewed as vectors of dimension $n^2$) where
\[
B:=\frac 1{nd}E-\frac 1{n^2}J.
\]
An upper bound $\delta$ on this product means that $G$ is a
$(K,\delta)$-partition expander.

We will use the Hoffman-Wielandt inequality (\theoref{t_HW}).
To this end we need to know the spectra of $A$ and
$B$.

The matrix $A$ has spectrum $(a_1,\dots,a_K,0,\dots,0)$,
because $\rank(A)\leq K$. Note that
\begin{equation}\label{e2}
\sum_{i=1}^K a_i^2=\|A\|_F^2\leq n^2.
\end{equation}
Let $d,\lambda_2,\dots,\lambda_n$ be the spectrum of $E$. The spectrum
of $J$ is $(n,0,\dots,0)$. The eigenspaces of $d$ and $n$ are the same
and all eigenspaces of $\lambda_i$, $i=2,\dots,n$ are in the
eigenspace of $0$ of $J$. Hence the spectrum of $B$ is
$\frac 1{nd}(0,\lambda_2,\dots,\lambda_n)$. 

Applying 
the Hoffman-Wielandt inequality we get
\[
(A,B)\leq 
\frac 1{nd}\max_\pi\sum_{i=1}^Ka_i\lambda_{\pi(i)}\leq
\frac 1{nd}\sum_{i=1}^K|a_i|\lambda_{2}\leq
\frac 1{nd} n\sqrt{K}\lambda_2=\sqrt{K}\lambda_2/d,
\]
where $\lambda_1=0$. The first inequality follows from the Hoffman-Wielandt
inequality in the form~\eqref{e-HW} and the last one follows from
\eqref{e2} and the Cauchy-Schwarz inequality.
\end{proof}

Now we will show that the bound proved above is essentially optimal, and therefore, in general expanders are not good partition expanders. 
We will use the following result of Alon and
Roichman~\cite{AR94_Ra}. (For a simpler proof, and an explicit and
better bound, see~\cite{LR04_Ra}.)

\begin{theorem}[\cite{AR94_Ra,LR04_Ra}]
There exists an absolute constant $c$ such that for every finite
group $\Gamma$ and any $d\leq|\Gamma|$, the following is true. If we
pick uniformly at random the elements $g_1,\dots,g_d\in \Gamma$, then
the resulting Cayley-graph has the second largest eigenvalue $\lambda$
satisfying
\[
\lambda\leq c\cdot\sqrt{d\log|\Gamma|}
\]
with probability going to $1$ as $|\Gamma|\to\infty$.
\end{theorem}
This theorem is not stated explicitly in those papers, but it is an
immediate corollary of Theorem~2 of \cite{LR04_Ra}. (One can take any
constant $c$ such that $c>2\ln 2$.)

Let $m>0$ be a natural number and let $\Gamma$ be the symmetric group on $m$
elements represented by permutations of $[m]$. Let 
$\pi_1,\dots,\pi_d$ be some permutations for which the
bound on the eigenvalue is satisfied. 
W.l.o.g. we will assume that for
every $i\in[d]$ there is a $j\in[d]$ such that $\pi_j=\pi_i^{-1}$. 
Let $G$ be the Cayley graph
determined by $\Gamma$ and  $\pi_1,\dots,\pi_d$. 

Let $1\leq t\leq m$. We will consider the partition $\{C_1\dc C_K\}$
defined by the following equivalence relation on $G$
\[
\rho|_{[t]}=\sigma|_{[t]},
\]
where $\rho,\sigma\in G$ are permutations and $|_{[t]}$ denote their
restriction to the first $t$ elements. 
Thus the number of blocks is $K=m(m-1)\dots(m-t+1)$. 
Consider the symmetric set $S$ defined by
\begin{equation}\label{e10}
(i,j)\in S\ \equiv\ \exists\rho\in C_i,\sigma\in C_j\ \exists s\in[d]\
\rho|_{[t]}=\pi_s\sigma|_{[t]}
\end{equation}
Note that if for some $i$ and $j$ the condition is satisfied by some
$s=s_0$, then for all $\rho\in C_i,\sigma\in C_j$, we have
$\rho|_{[t]}=\pi_{s_0}\sigma|_{[t]}$.

Consider the equation~\eqref{e1.5} that defines partition expanders. The
first term is in our case equal to 1. To bound the second term, note
that for a given $s\in[d]$ the number of pairs $\rho,\sigma$ satisfying the
condition $\rho|_{[t]}=\pi_\ell\sigma|_{[t]}$  is $m!(m-t)!$. Hence the second
term is bounded by
\[
\frac{d\cdot m!(m-t)!}{(m!)^2}=\frac d{m(m-1)\dots(m-t+1)}=\frac dK.
\]
This proves that if $d/K< 1-\delta$, then $G$ is not a
$(K,\delta)$-partition expander.

Thus we have proved:
\theo[t_HWtight]{
There exist a constant $c$ such that 
for infinitely many $n$ and every $d\leq n$, there are $d$-regular $c\sqrt{d\log n}$-expanders on $n$ vertices which are not $(K,1-\fr{d+1}K)$-partition expanders.
}

Comparing this statement to the bound given by \theoref{t_epe} in the most natural regime when a $(K,1-\asOm1)$-partition expander is required, we can see that the upper and the lower bounds match up to the factor of $\log n$ in the spectral gap.
In particular, since the second eigenvalue of a graph is always \asOm{\sqrt d}, $K$ can be at most linear in $d$, as long as our only assumption about $G$ is the absolute value of its second eigenvalue.
In contrast to this, according to Corollary~\ref{cr_r-reg}, there exist
$(K,1-\asOm1)$ partition expanders whose degree is $O(\log K)$. Thus any construction of such partition expanders must rely on some properties of $G$, other than the spectral gap.

\sect[s_SMP]{Partition expanders as PRGs in communication complexity}

Let us turn to the realm of communication complexity, where we give an equivalent formulation of partition expanders.
First, we use this equivalence to give a nearly-tight lower bound on the degree of good partition expanders, thus arguing near-optimality of the randomized construction given in \sref{s_rand}.
Second, we use the same construction to obtain a new separation between two models of communication complexity, which is qualitatively stronger than the previously known one.

We will use the following models of two-party communication
complexity.  

\ndefi[d_commode]{Models of communication complexity}{Two players whose names are Alice and Bob each receive a binary string of length $n$, respectively denoted by $x$ and $y$.
Players' goal is to compute the value of $f(x,y)$, where $f:\01^n\times\01^n\to\01$ is fixed.
The players obey the following scenario:\itemi{
\item In the model of \emph{Simultaneous Message Passing} (SMP), denoted by
\RII, both Alice and Bob send a message to the third participant,
the referee.  The referee does not know the values of $x$ and $y$,
so his only input are the messages received from the players, and
he has to produce the answer using the information received from
the players.  All three participants are allowed to use private
randomness.
\item The model of \emph{SMP with shared randomness,} denoted by \RIIp, is
similar to \RII\ but the players are allowed to use public
randomness.
\item In the model of \emph{One-Way Communication,} denoted by \RI, Alice
sends her message to Bob, who has to produce the answer using his
part of the input and the information received from Alice.
\item In the model of \emph{Simultaneous Two-Way Communication,} denoted by \Rz, Alice and
Bob send their messages simultaneously, similarly to the case of
SMP.  But here the recipient of Alice's message is Bob, and the
recipient of Bob's message is Alice.  Upon receiving the partner's
message, each player must produce an answer.}

We say that a communication protocol solves the problem represented by $f$ if it produces the correct answer(s) with probability at least $\dr23$ for every possible input.
The communication cost of a protocol is the maximal total number of bits sent by the players, and the communication cost of a function $f$ is the minimal communication cost of a protocol that solves it in the given model.}

The models \RII, \RIIp\ and \RI\ have been studied widely and the corresponding notation is commonly used; the Simultaneous Two-Way model has been considered in several works (see below), but no specific name was assigned to it.
Note that when we say that an \Rz-protocol has produced the answer ``$a$'', we refer to the situation when both the players have produced the same answer.  

\ndefi[d_pr-com]{Pseudo-randomness in communication complexity}
{Let $\Cl M$ be a communication complexity model, and let $\mu$ be a distribution defined over $\01^n\times\01^n$.
We say that $\mu$ is \f k-pseudo-random for $\Cl M$ if for any protocol $\Cl P$ of communication cost at most $k$ it holds that
\m{\PR[(X,Y)\sim\mu]{\Cl P(X,Y)\txt{ outputs ``$1$''}}
-\PR[{(X,Y)\sim\U[\01^n\times\01^n]}]
{\Cl P(X,Y)\txt{ outputs ``$1$''}}<\fr13.}

We say that $g:\01^s\to\01^n\times\01^n$ is a \f k-Pseudo-Random Generator (\f k-PRG) of seed length $s$ against $\Cl M$ if the distribution of $g(X)$ when $X\sim\U[\01^s]$ is \f k-pseudo-random for $\Cl M$.
}

Pseudo-randomness in the context of communication complexity has been introduced in \cite{INW94_Pse}.
Intuitively, both pseudo-randomness and lower bounds on communication cost can be viewed as claims that certain problem is hard for the model under consideration.

Given a \f d-regular graph $G=(\01^n,E)$, let $\mu_G$ be the uniformly random distribution of the edges from $E$.
Note that in order to choose $(v_1,v_2)\sim\mu_G$, a ``seed'' of
length $n+\log d$ is both necessary and sufficient.

\theo[t_equi2]{Let $k,n\in\NN$ and $G=(\01^n,E)$.
The following statements are equivalent:\enum{
\item $G$ is a $(2^{\asT k},\delta)$-partition expander for some $\delta<\dr13$.
\item $\mu_G$ is \asT k-pseudo-random for \RII.
}}

In particular, our construction in \sref{s_rand} corresponds to a \f k-PRG against \RII\ of seed length $n+\asO{\log k}$.
Note that due to the fact that in the context of communication complexity the players are computationally unlimited, a randomized construction of a PRG is neither meaningless nor trivial.\fn
{For example, the models \RI\ and \Rz\ (and more generally, any two-party model where a \f k-bit message from one player reaches the other player, who also receives his own $n$ bits of input) require seed length at least $n+k-\asO1$ even with a non-uniform PRG, as witnessed by the protocol where the sender sends the first \f k\ bits of his input and the computationally-unlimited recipient outputs ``$1$'' only if the message together with his own $n$ bits of input have Kolmogorov complexity $n+k-\asO1$.}
Note also that a distribution is pseudo-random against \RII\ if and only if it is pseudo-random against \RIIp, as for any distribution-distinguishing task there exists an optimal protocol that does not use randomness.

\prf{Let $C$ be a constant.
First, suppose that $G$ is a $(2^{Ck},\delta)$-partition expander.
Let $\Cl P$ be an \RII-protocol of cost at most $Ck$, and let us show that it cannot distinguish with high confidence $\mu_G$ from $\U[\01^n\times\01^n]$.
Without loss of generality assume that $\Cl P$ is deterministic, and let $\alpha:\01^n\to\01^{Ck}$ be the mapping from $x$ to the concatenation of Alice's and Bob's messages in response to the input $(x,x)$.
Let $\nu_{\U}$ and $\nu_G$ be the distributions of $(\alpha(X),\alpha(Y))$ when, respectively, $(X,Y)\sim\U[\01^n\times\01^n]$ and $(X,Y)\sim\mu_G$.
Clearly,
\m{\PR[(X,Y)\sim\mu_G]{\Cl P(X,Y)\txt{ outputs ``$1$''}}
-\PR[{(X,Y)\sim\U[\01^n\times\01^n]}]
{\Cl P(X,Y)\txt{ outputs ``$1$''}}
\le\dst(\nu_G,\nu_{\U}).}
Note that $\alpha$ defines a partition of $\01^n$ into at most $2^{Ck}$ blocks, and by the definition of partition expanders, 
\m{\dst(\nu_G,\nu_{\U})\le\delta<\dr13.}
Therefore, $\mu_G$ ``fools'' $\Cl P$ and thus it is $Ck$-pseudo-random for \RII.

Now assume that $\mu_G$ is $2Ck$-pseudo-random for \RII, and let us show that $G$ is a partition expander.
Let $\sigma=\set{S_1\dc S_{2^{Ck}}}$ be a partition of $\01^n$, and
for $x\in\01^n$, define $\sigma(x)\deq i$ for $i$ such that $x\in S_i$.
Let $\Cl P_\sigma$ be an \RII-protocol, where upon receiving input $(X,Y)$, Alice sends $\sigma(X)$ and Bob sends $\sigma(Y)$.
Let $\tau_{\U}$ and $\tau_G$ be the distributions of $(\sigma(X),\sigma(Y))$ when, respectively, $(X,Y)\sim\U[\01^n\times\01^n]$ and $(X,Y)\sim\mu_G$.
Note that $\Cl P_\sigma$ is of cost $2Ck$, and therefore
\m{\dst(\tau_{\U},\tau_G)<\dr13,}
since otherwise the referee would be able to distinguish the two cases with confidence high enough to contradict pseudo-randomness of $\mu_G$.
Let $\delta$ be the maximum value of $\dst(\tau_{\U},\tau_G)$ possible under any choice of $2^{Ck}$-partition $\sigma$, then $\delta<\dr13$ and $G$ is a $(2^{Ck},\delta)$-partition expander, as required.}

\ssect[s_low]{Lower bound on the degree of partition expanders}

Let us use the correspondence between partition expanders and pseudo-random generators given by \theoref{t_equi2} in order to get a lower bound on the degree of partition expanders.

\theo[t_dlow]{For any $\delta<\dr13$, if a \f d-regular graph $G$ is a $(K,\delta)$-partition expander then $d\in\asOm{\fr{\log K}{\log\log K}}$.}

In particular, the randomized construction given in \sref{s_rand} is optimal, up to the multiplicative $\log\log K$ factor.

\prf{For convenience, let $n$ and $d$ be powers of $2$.
Let $G=([n],E)$, and assume it is a $(K,\delta)$-partition expander.
On the one hand, according to \theoref{t_equi2}, $\mu_G$ is \asOm{\log K}-pseudo-random for the SMP model.
On the other hand, we will see below that an SMP protocol of cost \asO{d\log d} can distinguish $\mu_G$ from the uniform distribution with error at most \dr 14, and therefore $d\in\asOm{\fr{\log K}{\log\log K}}$, as required.

The distinguishing protocol is as follows.
When her input is $v\in V$, Alice sends to the referee the first $\log d+2$ bits of the indices of the $d$ neighbors of $v$.
On input $u\in V$, Bob sends to the referee the first $\log d+2$ bits of the index of $u$.
The referee guesses that the input pair $(v,u)$ has been drawn from the distribution $\mu_G$ if the index-prefix received from Bob appears in the list of $d$ index-prefixes received from Alice.
This protocol is always correct if the input comes from the support of
$\mu_G$, and errs with probability at most \dr14\ when the input comes
from the uniform distribution.}

\ssect{Model separations based on PRGs}

Model separation in computational complexity usually means demonstrating existence of a computational problem that can be solved efficiently in one model, but not in the other.
If several classes of problems can be handled by the models under consideration, one can define the corresponding \e{types} of model separations.
When one problem class is a special case of another, separation via an element of the smaller class can be viewed as a stronger indication that the compared models have different computational power than separation via an element of the bigger class.
These ideas can be pushed further, resulting in various ``hierarchies'' of model separations.

In the case of communication complexity, there are at least four natural classes of computational problems\fn
{The same applies to many other fields of complexity, where also most of the following discussion remains valid -- e.g., in the field of circuit complexity.},
namely:\itemi{
\item Total functions $f:A\times B\to Z$
\item Partial functions $f:C\to Z$, $C\sbseq A\times B$
\item Relations $\Cl P\sbseq A\times B\times Z$
\item Distinguishing some distribution $\mu$ defined on $A\times B$ from the uniform (cf.~\defiref{d_pr-com})
}

Consider the four types of model separations corresponding to these
four classes. We will call the fourth type \e{separation via a PRG}.
Obviously, if two communication models are separable via a total function they are also separated via a partial function, and separability via a partial function implies separability via a relation.
On the other hand, there are pairs of communication models that can be separated via a relation but not via a partial function (e.g., see~\cite{GRW08_Sim_Co}), and there are many pairs of models that have been separated via partial functions, but are \e{conjectured} not to be separable via total functions (e.g., most of quantum communication models form such pairs with their natural classical counterparts).
Therefore, in communication complexity it is always desirable to separate models via the ``most limited'' possible type of separation, as that gives the ``strongest'' possible indication of difference in the computational power of those models.

To the best of our knowledge, separation via a PRG has not been studied in the context of communication complexity.
It is probably incomparable to the first three types of separation:
On the one hand, it is straightforward to get a separation via a PRG by modifying slightly one of the known separations via a partial function between quantum and classical one-way models, but it is \e{conjectured} that those two models cannot be separated via a total function.
Therefore, modulo that conjecture, separation via a PRG cannot, in general, be as limited as separation via a total function.
On the other hand, the models \RII\ and \RIIp\ cannot be separated via a PRG (because for any distribution-distinguishing task there exists an optimal protocol that does not use randomness), but they can be separated via a total function -- e.g., the equality function.
Therefore, separation via a total function cannot, in general, be as limited as separation via a PRG.

Is there a type of model separation that would be the most limited, and therefore separations demonstrated through it would be the most ``convincing'' indication of difference in the computational power of the compared models?

Take a total Boolean function $f:A\times B\to\01$, let $\Cl M$ be a communication complexity model, and consider the following two claims:\itemi{
\item No protocol in $\Cl M$ of cost less than $k$ can compute $f$.
\item The distributions $\U[f^{-1}(0)]$ and $\U[f^{-1}(1)]$ are \f k-PRGs for $\Cl M$.}
We will say that $f$ is \e{\f k-hard for $\Cl M$} in the first case, and that $f$ is \e{\f k-pseudo-random for $\Cl M$} in the second.\fn
{Note that we required both $\U[f^{-1}(0)]$ and $\U[f^{-1}(1)]$ to be \f k-PRGs for $\Cl M$ when $f$ is \f k-pseudo-random in order not to require $f$ to be balanced; if it is balanced, either condition implies the other.}
If $f$ is \f k-pseudo-random for $\Cl M$, then it is also \f k-hard for $\Cl M$; the converse is not necessarily true, as follows from the same example of the equality function in \RII.

As usual in communication complexity, we will say that a communication problem is \e{easy} for a given model if it can be solved by a protocol of cost $(\log n)^{\asO1}$.

\ndefi{Ultra-separation}{Complexity models $\Cl M_1$ and $\Cl M_2$ are ultra-separated if there is a total Boolean function $f$ that is easy for $\Cl M_1$ and $n^{\asOm1}$-pseudo-random for $\Cl M_2$.}

Ultra-separation is a very limited type of model separation -- in fact, the most limited ``reasonable'' one we came up with.

\clm{For any two models that allow efficient error reduction for total functions, ultra-separability implies separability both via a total function and via a PRG.}

Here by efficient error reduction we mean that if $f$ can be solved efficiently, then for any constant $\eps$ there exists an efficient protocol that solves $f$ with error at most $\eps$.
Probably all studied communication complexity models satisfy this very natural property.

\prf{If $f$ is $n^{\asOm1}$-pseudo-random for $\Cl M_2$, then it is also $n^{\asOm1}$-hard for $\Cl M_2$, and therefore ultra-separability implies separability via a total function.

If $f$ is easy for $\Cl M_1$, then the elements of $f^{-1}(1)$ can be distinguished from the elements of $f^{-1}(0)$ with worst-case error at most $\dr1{10}$ by a protocol of cost $(\log n)^{\asO1}$.
Without loss of generality, let $\PR{f(X,Y)=1}\le\dr12$ when $(X,Y)$ is uniformly random.
Then there exist an efficient protocol in $\Cl M_1$ that outputs ``$1$'' with probability at least $\dr9{10}$ when $(X,Y)\sim\U[f^{-1}(1)]$, and with probability at most $\dr{11}{20}$ when $(X,Y)$ is uniformly random.
So, $\U[f^{-1}(1)]$ can be distinguished from the uniform with ``bias'' more than $\dr13$ by an efficient protocol in $\Cl M_1$, and thus it is not a PRG.
Therefore, ultra-separability implies separability via a PRG.}

\ssect{Ultra-separation of \RIIp\ and \Rz}

We have seen that \e{ultra-separability of two models is a stronger evidence of difference in their computational power than separability via a function (total or partial), via a relation, or via a PRG}.
We are not aware of any type of model separation that would not be subsumed by ultra-separation.
Therefore, it is interesting to demonstrate ultra-separations even for those pairs of models that have been separated previously via some ``less convincing'' methods.

For long time, it had been believed that the models \RIIp\ and \Rz\ were equivalent.
In 2002 Bar-Yossef, Jayram, Kumar and Sivakumar~\cite{BJKS02_In_Th} demonstrated a separation between these models via a cleverly constructed total function $g$, for which $\Rz(g)\in\asO{\log n}$ and $\RIIp(g)\in\asOm{\sqrt n}$.
The ideas used in their construction seem to be insufficient to yield separation via a PRG.

\theo[t_sep]{The models \RIIp\ and \Rz\ can be ultra-separated.
Namely, there exists a total Boolean function $f$, such that $\Rz(f)\in\asO{\log n}$ and $\U[f^{-1}(1)]$ cannot be distinguished from $\U[\01^n\times\01^n]$ by any \RIIp-protocol of cost \aso n.}

The new separation is stronger not only qualitatively, but quantitatively as well -- the improvement results from the (optimal) lower bound of \asOm{n} on the \RIIp-complexity of $f$.

\prf{From \crlref{cr_r-reg} it follows that for any constant $\delta$ there exists a graph $G$ on $2^n$ vertices of degree $d\in\asT n$, which is a $(2^{\dr n2},\delta)$-partition expander.
According to \theoref{t_equi2}, the corresponding $\mu_G$ is \asT n-pseudo-random for \RII.
Clearly, the same is true for $\mu_{\bar G}$, where $\bar G$ is the complement graph.
If we define $f_G:\01^n\times\01^n\to\01$ to be the ``edge function'' of $G$, then it is \asT n-pseudo random for \RII.

Let us see that $\Rz(f_G)\in\asO{\log d}=\asO{\log n}$.
Consider a protocol where the players use shared randomness\fn
{The power of the model \Rz\ is not affected by allowing public randomness.}
to choose a hash function from $\01^n$ to $\01^{2\log d}$, then Alice sends the hash-value of $x$ and Bob answers ``$1$'' if the received value equals the hash-value of one of the neighbors of $y$ in $G$, and ``$0$'' otherwise (if Bob is the sender, they act symmetrically).
This protocol has communication cost \asO d\ and computes $f_G$ with error \aso1.
The result follows.}

\sect{Discussion}

The most interesting open problem is to find an explicit construction
of a good partition expander; more precisely, to construct a family of
$(K,\delta)$-partition expanders in which $\delta<1$ is constant, $K$
goes to infinity, and the degrees are $d=O(\log K)$. We will call
informally such families of graphs \emph{good partition expanders}.
As we have shown in this paper, expanders are, in general, not good
partition expanders and it seems unlikely that the property would be
implied by a property of the spectrum of a graph. One possible way
of constructing good partition expanders could be by using zig-zag
product or a similar kind of product. Indeed, in a recent paper
\cite{MN13_Non} Mendel and Naor have shown that zig-zag product can
be used for constructing various types of generalizations of
expanders. These constructions start with a small object, which can be
found by brute force, and which are enlarged by applying products
repeatedly. They work well when one needs constant degree, but in our case
we need increasing degree and to satisfy a certain property for
partitions with exponentially increasing number of blocks. It is not
totally excluded that some kind of product will work, but it will
require a new kind of argument to prove it.

We demonstrated some applications of partition expanders in communication complexity.
In particular, we defined the notion of ultra-separation and argued that it is one of the weakest model-separating methods, thus applying it provides a very strong (probably, the strongest known) evidence that the two separated models have different computational power.
We gave an example of such separation.
It would be interesting to find more examples of ultra-separations, not only in communication complexity.

We believe that partition expanders will be useful in many other areas of complexity theory, especially when explicit constructions are found.
For example, one could use good partition expanders instead of expanders in the pseudorandom generators of Impagliazzo, Nisan and Wigderson~\cite{INW94_Pse}, provided that an explicit construction of good partition expanders is found.
Since the number of partitions corresponds to the exponential of space complexity, they would certainly have better parameters.
This, however, requires further research, because the direct application of partition expanders in INW generators does not seem to give substantially better results than the use of expanders.

\section*{Acknowledgments}

We thank Hartmut Klauck and anonymous reviewers for helpful comments.

\bibliography{tex}

\begin{thebibliography}{BYJKS02}

\bibitem[AR94]{AR94_Ra}
N.~Alon and Y.~Roichman.
\newblock {Random Cayley Graphs and Expanders}.
\newblock {\em Random Structures and Algorithms 5}, pages 271--284, 1994.

\bibitem[AS08]{AS08_The}
N.~Alon and J.~Spencer.
\newblock {The Probabilistic Method}.
\newblock {\em John Wiley}, 2008.

\bibitem[Azu67]{A67_We}
K.~Azuma.
\newblock {Weighted Sums of Certain Dependent Random Variables}.
\newblock {\em Tohoku Mathematical Journal 68}, pages 357--367, 1967.

\bibitem[Bol80]{B80_A_Pro}
B.~Bollob\'as.
\newblock {A Probabilistic Proof of an Asymptotic Formula for the Number of
  Labelled Regular Graphs}.
\newblock {\em European Journal of Combinatorics 1}, pages 311--316, 1980.

\bibitem[BYJKS02]{BJKS02_In_Th}
Z.~Bar-Yossef, T.~S. Jayram, R.~Kumar, and D.~Sivakumar.
\newblock {Information Theory Methods in Communication Complexity}.
\newblock {\em Proceedings of 17th IEEE Conference on Computational
  Complexity}, pages 93--102, 2002.

\bibitem[CRVW02]{CRVW02_Ran}
M.~Capalbo, O.~Reingold, S.~Vadhan, and A.~Wigderson.
\newblock {Randomness Conductors and Constant-Degree Lossless Expanders}.
\newblock {\em Proceedings of the 34th Symposium on Theory of Computing}, pages
  659--668, 2002.

\bibitem[DW10]{DW10_Mon}
Z.~Dvir and A.~Wigderson.
\newblock {Monotone Expanders: Constructions and Applications}.
\newblock {\em Theory of Computing 6(1)}, pages 291--308, 2010.

\bibitem[GRdW08]{GRW08_Sim_Co}
D.~Gavinsky, O.~Regev, and R.~de~Wolf.
\newblock {Simultaneous Communication Protocols with Quantum and Classical
  Messages}.
\newblock {\em Chicago Journal of Theoretical Computer Science}, 7, 2008.

\bibitem[HW53]{HW53_The}
A.~J. Hoffman and H.~W. Wielandt.
\newblock {The Variation of the Spectrum of a Normal Matrix}.
\newblock {\em Duke Mathematical Journal 20}, pages 37--39, 1953.

\bibitem[INW94]{INW94_Pse}
R.~Impagliazzo, N.~Nisan, and A.~Wigderson.
\newblock {Pseudorandomness for Network Algorithms}.
\newblock {\em Proceedings of the 26th Symposium on Theory of Computing}, pages
  356--364, 1994.

\bibitem[LR04]{LR04_Ra}
Z.~Landau and A.~Russell.
\newblock {Random Cayley Graphs are Expanders: A Simple Proof of the
  Alon-Roichman Theorem}.
\newblock {\em Electronic Journal of Combinatorics 11}, 2004.

\bibitem[MN13]{MN13_Non}
M.~Mendel and A.~Naor.
\newblock {Nonlinear Spectral Calculus and Super-Expanders}.
\newblock {\em Publications math\'ematiques de l'IH\'ES}, 2013.

\bibitem[MW91]{KW91_Asy}
B.~D. McKay and N.~C. Wormald.
\newblock {Asymptotic Enumeration by Degree Sequence of Graphs with Degrees
  $o(n^{1/2})$}.
\newblock {\em Combinatorica 11(4)}, pages 369--382, 1991.

\bibitem[Wor99]{W99_Mo}
N.~C. Wormald.
\newblock {Models of Random Regular Graphs}.
\newblock {\em Surveys in Combinatorics. Lecture Note Series 276}, pages
  239--298, 1999.

\end{thebibliography}

\MyComment{Look for ...-s}

\MyComment{Spell-check}

\end{document}